\documentclass[a4paper,fleqn]{cas-dc}

\usepackage[numbers]{natbib}

\usepackage{amsmath}
\DeclareMathOperator*{\argmin}{argmin}
\def\tsc#1{\csdef{#1}{\textsc{\lowercase{#1}}\xspace}}
\tsc{WGM}
\tsc{QE}
\tsc{EP}
\tsc{PMS}
\tsc{BEC}
\tsc{DE}

\begin{document}
\sloppy
\let\WriteBookmarks\relax
\def\floatpagepagefraction{1}
\def\textpagefraction{.001}
\shorttitle{Lensless microscopy with light source of low spatio-temporal coherence}
\shortauthors{S Kumar et~al.}

\title [mode = title]{Lensless in-line holographic microscopy with light source of low spatio-temporal coherence}                      



\author[1]{Sanjeev Kumar}[orcid=0000-0002-9943-6343]
\cormark[1]
\fnmark[1]
\ead{sanjeevsmst@iitkgp.ac.in}


\address[1]{School of Medical Science and Technology, Indian Institute of Technology, Kharagpur, 721302, India}

\author[1]{Manjunatha Mahadevappa}[]

\author[2]{Pranab K. Dutta}[]

\address[2]{Department of Electrical Engineering, Indian Institute of Technology, Kharagpur, 721302, India}

\cortext[cor1]{Corresponding author}


\begin{abstract}
Lensless microscopy with coherent or partially coherent light sources is a well known imaging technique, commonly referred as digital in-line holographic microscopy. In the established methods, both the spatial and temporal coherence of light play a crucial role in determining the resolution of reconstructed object. We report lensless microscopy with a spatially extended white LED, a light source of low spatial and very low temporal coherence. The wave-field propagation between two parallel planes can be obtained using a convolution operation, where the convolution kernel depends on the object-sensor distance and the characteristics of the light source. For a light source of unknown characteristics, this kernel is an unknown function. In the proposed reconstruction method, we decompose an unknown convolution kernel of very large size ($128 \times 128$) into a small unknown light-source-specific kernel (size $9 \times 9$) and a known light-source-independent kernel (size $128 \times 128$). This drastically reduces the number of parameters to be estimated at the system identification step, which has been performed here by one time imaging of the known microscopic objects. Final unknown object estimation has been performed using the upper-bound constrained deconvolution. A lateral resolution of $\sim$1-2\ $\mu$m has been demonstrated.
\end{abstract}



\begin{keywords}
Digital in-line holographic microscopy \sep Lensless microscopy \sep Computational imaging \sep Image reconstruction
\end{keywords}

\maketitle

\section{Introduction}

Lensless microscopy has gained its popularity due to its simple, compact, portable and cost-effective hardware. Large field of view images at the sub-micron resolution can be obtained \cite{ozcan2016lensless}. It has been widely investigated for imaging of weakly scattering objects and commonly referred as digital in-line holographic microscopy. In single-shot lensless microscopy methods, the hardware consists of a light source, a sample holder, an image sensor and an image processing system (see figure 1). In the established methods, the high spatial and temporal coherence of light play a crucial role in determining the high resolution of reconstructed object \cite{agbana2017aliasing,hagemann2018coherence}. Various authors have used monochromatic laser or narrow band LED ($\Delta \lambda = \sim 20-25 $ nm) coupled to a single mode optical fiber or a pinhole/lens-pinhole filtering system \cite{garcia2016color}.

Recorded fringe pattern on the image sensor is optionally passed through a denoising \cite{yan2018fringe} or a fringe enhancement step \cite{greenbaum2013increased,byeon2019deep}. The principle of image recording is based on Gabor in-line holography \cite{gabor1948new}, where the recorded intensity at the sensor plane $I(\textbf{x})$ is expressed as the interference between the unscattered wave-field $A$ and the scattered wave-field $U(\textbf{x})$ \cite{goodman2005introduction}:
\begin{align}
I(\textbf{x}) &= |A|^2 + |U(\textbf{x})|^2 + A^*U(\textbf{x}) + A U^*(\textbf{x}) \\
&=  A^*U(\textbf{x}) + e
\end{align}
where $\textbf{x}=(x,y)$ is the lateral coordinate vector and $e$ denotes the unwanted terms in holography. Wave-field propagation between two parallel planes is obtained by solving scalar diffraction integral equations such as Rayleigh- Sommerfeld diffraction formula or Fresnel-kirchoff diffraction formula \cite{born2013principles, goodman2005introduction}. The field at observation plane $U(\textbf{x})$ is expressed as the convolution of object $o(\textbf{x})$ with the impulse response function $h(\textbf{x})$ given by these equations. First solution of Rayleigh-Sommerfeld formula gives:
\begin{equation}
U(\textbf{x}) = h(\textbf{x}) \otimes o(\textbf{x});\ \text{where} \
h(\textbf{x}) = \frac{1}{j\lambda} \frac{exp(jk_0r)}{r} cos \ \phi
\end{equation}
where $\otimes$ denotes the convolution operator, $\lambda$ is the wavelength, $k_0 = \frac{2 \pi}{\lambda}$ is the wave-number, $r = \sqrt{x^2 + y^2 + z^2}$, $z$ is the distance from the object to observation plane and $cos \ \phi$ is called the obliquity factor. Here $\phi$ is the angle between the unit normal vector $n$ and the distance vector $r$. Under the paraxial approximation, $cos \ \phi \approx 1$ \cite{goodman2005introduction}. Alternatively, angular spectrum method is used for the wave-field propagation between two parallel planes which involves computing the fourier transform of object, multiplication with the free space optical transfer function $H_{OTF}$ and computing the inverse fourier transform \cite{matsushima2009band, kozacki2012computation}. The free space optical tranfer function is obtained by the following equation (for refractive index, $n=1$ for free space) \cite{ozcan2016lensless}:
\begin{align}
&H_{OTF}(\textbf{v}) =  exp\Big(jk_0z \sqrt{1 - (\lambda v_x)^2-(\lambda v_y)^2}\Big); \\ &v^2_x + v^2_y < \frac{1}{\lambda^2} \\
&H_{OTF}(\textbf{v}) =  0; \ v^2_x + v^2_y \geq \frac{1}{\lambda^2} 
\end{align}
where, $\textbf{v}=(v_x,v_y)$ is the frequency coordinate vector. Wave-field backpropagation is obtained as \cite{ozcan2016lensless}:
\begin{equation}
\text{Reconstructed object} = IFT \Big[ H^{-1}_{OTF}(\textbf{v}) \ FT \big[I(\textbf{x})\big] \Big]
\end{equation}
where $IFT$ and $FT$ denote the inverse fourier transform and fourier transform respectively. Since it is difficult to accurately record the value of $z$ while capturing fringe patterns, an iterative assessment is required to estimate the value of z which provides the sharpest object reconstruction.
Finally, an iterative phase retrieval step is used to suppress the twin image artifact (ringing artifact around the features) contributed by the conjugate scattered field $AU^*(\textbf{x})$ in equation 1. Either some prior information about the object \cite{fienup1982phase,mudanyali2010compact} or multiple intensity recordings are used at this step \cite{bao2008phase,maiden2009improved, mahajan2016wide}. Recently, deep learning and optimization based algorithms have been used to suppress these artifacts \cite{rivenson2018phase,jolivet2018regularized}.

Limited pixel-pitch of the sensor, signal-to-noise ratio and the coherence of light are major factors that influence the resolution of the reconstructed object. Nowadays, image sensors with a pixel-pitch $\sim1 \ \mu$m  are commercially available. Signal-to-noise ratio is improved by using the statistical methods of object reconstruction, such as optimization of an appropriate cost function \cite{bertero1998introduction, de2016limits}. Cost function includes a data fidelity term, a regularization function term and an empirically decided regularization parameter in most cases. Regularization part is carefully designed functional of object based on the prior information (about the object) essentially to prevent noise overfitting.
\begin{equation}
E(\hat{o}) = ||g - f(\hat{o})|| + \gamma \ r(\hat{o})
\end{equation}
Here $E$ is the cost function, $\hat{o}$ is an estimate of object, $\lVert(.)\rVert$ denotes $l-2$ norm, $g$ is the recorded fringe pattern, $f(.)$ is the function to transform object into fringe pattern, commonly referred as the known forward problem, $\gamma$ is the regularization parameter and $r(.)$ is the regularization function.

In this paper, we focus our main discussion on the resolution deteriorated due to the reduced coherence of the light source. Low spatial and/or temporal coherence lead to observation of highly blurred fringes with reduced contrast and hence loss of resolution in the final reconstructed image. In other words, it reduces the limit of the maximum fringe frequency which can be detected (at a given working distance), which, in turn, limits the maximum resolution in the reconstructed object \cite{agbana2017aliasing}. On the other hand, when highly coherent sources are used, reconstructed image is corrupted with the speckle noise and ringing artifacts due to the interference of light reflected at several layers between the sensor and the source \cite{goodman2007speckle, kanka2011high}.

\cite{feng2017resolution} used a spatially extended, narrow band LED (center wavelength = 470 nm) in lensless microscopy and demonstrated that the resolution reduced due to low spatial coherence can be improved by an additional deconvolution step before image reconstruction. They reported a lateral resolution of 3.1 $\mu$m. \cite{feng2019differential} used sunlight as the light for lensless microscopy and reported a resolution of 3.48 $\mu$m with their differential holographic reconstruction method. In another reconstruction method \cite{feng2019color}, they used the spectral distribution of sunlight and spectral response of sensor to obtain a synthetic point spread function, which they used to reconstruct object using wiener deconvolution. They reported a resolution of $\sim$6 $\mu$m with this method. 

Previously we reported lensless microscopy with a light source of low spatial and very low temporal coherence, a spatially extended white LED. We demonstrated a reconstruction resolution of $\sim$2 $\mu$m with suppressed ringing artifacts using a constrained and regularized optimization method \cite{kumar2019extended}. In this paper, we obtain lensless microscopy with the same light source and propose an algorithm for high resolution image reconstruction. This method involves one time imaging of known microscopic objects to estimate a small light-source-specific kernel. This kernel depends on the spatial and temporal coherence of the light source, image sensor's spectral response function and the positions of the source, sample and sensor. In other words, this kernel is highly specific to the design and components of a lensless imaging system. This kernel along with the optical transfer function for coherent light has been used to estimate high resolution images of the unknown microscopic objects in the proposed algorithm (see figure 2).

 \begin{figure}
\centering
{\includegraphics[width=6cm]{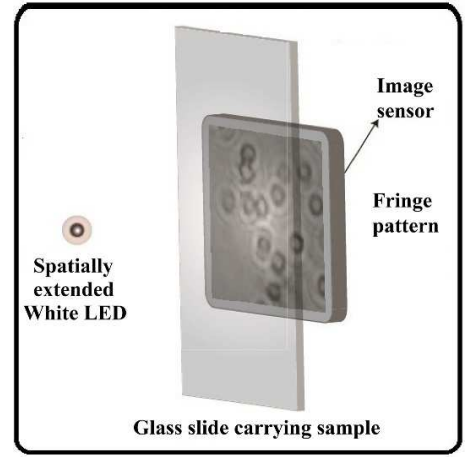}}
\caption{Optical setup for lensless microscopy with a spatially extended white LED.}
\label{Fig:1}
\end{figure}

 \begin{figure*}
\centering
{\includegraphics[width=12cm]{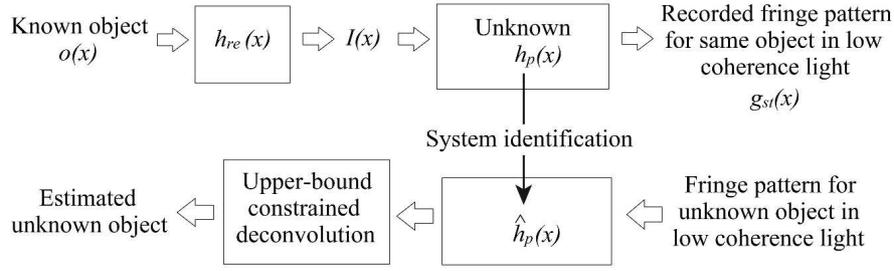}}
\caption{Proposed principle for lensless microscopy with a light source of low spatio-temporal coherence.}
\label{Fig:2}
\end{figure*}

\section{Methods} 
\subsection{In-line holography}
For the weakly scattering objects, $|U(\textbf{x})| << |A|$ and the equation 1 takes the following form:
\begin{align}
I(\textbf{x}) &\approx |A|^2 + A^*U(\textbf{x}) + \big(A^*U(\textbf{x})\big)^* \\
&=  \operatorname{Re}\big[|A|^2\big] + 2\operatorname{Re}\big[A^*U(\textbf{x}) \big] \\ &= \operatorname{Re}\big[|A|^2 + 2A^*U(\textbf{x}) \big] \\
&= \operatorname{Re}\Big[|A|^2 + 2A^*\Big(h(\textbf{x}) \otimes  o(\textbf{x})\Big)\Big] 
\end{align}

Now, we show that $|A|^2 = h(x) \otimes \frac{1}{c}|A|^2$, where $c = \int_{-\infty}^{+\infty}h(\textbf{x})dx$.
\begin{align}
h(\textbf{x}) \otimes \frac{1}{c}|A|^2 &= \frac{1}{c}|A|^2\Big(h(\textbf{x}) \otimes l(\textbf{x})\Big)\\ &= \frac{1}{c}|A|^2\int_{-\infty}^{+\infty}h(\textbf{x})dx = |A|^2
\end{align}
where $l(\textbf{x}) =$ all ones. 
Now the equation 12 becomes,
\begin{align}
I(\textbf{x})&=\operatorname{Re}\Big[\Big(h(\textbf{x}) \otimes \frac{1}{c}|A|^2 \Big) + \Big(h(\textbf{x}) \otimes 2A^* o(\textbf{x})\Big)\Big] \\ &=\operatorname{Re}\Big[h(\textbf{x}) \otimes \Big(\frac{1}{c}|A|^2  +  2A^* o(\textbf{x})\Big)\Big] 
\\ &=\operatorname{Re}\Big[h(\textbf{x}) \otimes f(\textbf{x}) \Big]
\end{align}
where $\ f(\textbf{x}) =  \frac{1}{c}|A|^2  +  2A^* o(\textbf{x})$ is the scaled form of the true object $o(\textbf{x})$. From equation 17 we get:
\begin{equation}
I(\textbf{x})=  \big[ h_{re}(\textbf{x}) \otimes f_{re}(\textbf{x})\ \big]  \\  -    \big[ h_{im}(\textbf{x}) \otimes f_{im}(\textbf{x})\  \big] 
\end{equation}
subscripts $(.)_{re}$ and $(.)_{im}$ denote the real and imaginary parts of the respective quantity $(.)$.
We assume that the phase shift introduced by our object is negligible i.e. $f_{im}(\textbf{x})=0$ and $f(\textbf{x})=f_{re}(\textbf{x})$. Now the equation 18 becomes:
\begin{equation}
I(\textbf{x}) \approx   h_{re}(\textbf{x}) \otimes f(\textbf{x})\ ) 
\end{equation}

\begin{figure}
\centering
{\includegraphics[width=8cm]{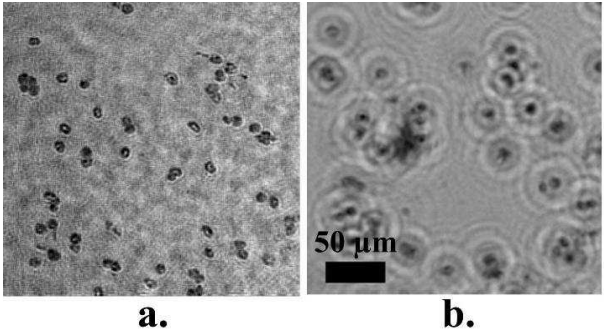}}
\caption{Red blood cells of diameter $\sim$6-8 $\mu$m reconstructed with angular spectrum method in the lensless microscopy setup (a.) when a monochromatic laser coupled to a single mode optical fiber is used as the light source; (b.) when a spatially extended white LED is used as light source. No phase retrieval algorithm or any additional image processing is applied in any of the two cases.}
\label{Fig:3}
\end{figure}

\subsection{Imaging in low coherence}
In the previously established holographic in-line microscopy methods, the forward problem is designed on the assumption that a monochromatic point source of light at a finite or infinite distance (in the case a collimator lens is used) is illuminating the sample. A spatially extended light source can be considered to contain many point sources, which are incoherent to each other. The observed fringe pattern in such a case is the superposition of individual fringe patterns produced by each point source, weighted by the scale factor $w_s$, proportional to the intensities of these sources. 
 \begin{equation}
 g_s(\textbf{x}) = \sum^m_{s=1} w_s \big[ h_s(\textbf{x}) \otimes f(\textbf{x}) \big] 
 \end{equation}
where $h_s$ denotes the real part of impulse response function for $s$th monochromatic point source (based on equation 19). Similarly, in case of a polychromatic light source, observed fringe pattern can be approximated as the superposition of fringe patterns produced by discrete wavelengths, weighted by the spectral distribution function $w_{\lambda}$  \cite{leith1977white, goodman2015statistical}.
  \begin{equation}
 g_{st}(\textbf{x}) = \sum^n_{\lambda=1} w_{\lambda} \Big[ \sum^m_{s=1} w_s \big[ h_{s \lambda}(\textbf{x}) \otimes f(\textbf{x}) \big] \Big] 
 \end{equation}
 where $h_{s \lambda}$ denotes the real part impulse response function for $s$th monochromatic point source of wavelength $\lambda$.

Using the distributivity property of convolution, equation 21 can be re-expressed with a single convolution operation as:
  \begin{align}
 g_{st}(\textbf{x}) &= \Bigg[ \Big[ \sum^n_{\lambda=1} w_{\lambda}  \sum^m_{s=1} w_s  h_{s \lambda}(\textbf{x}) \Big] \otimes f(\textbf{x})  \Bigg] \\
 &= \big[ h_{st}(\textbf{x}) \otimes f(\textbf{x}) \big] \\
\textnormal{where} \ &h_{st}(\textbf{x})=\sum^n_{\lambda=1} w_{\lambda}  \sum^m_{s=1} w_s  h_{s \lambda}(\textbf{x}) 
 \end{align}
Note that in this paper, we consider that the properties of the light source are not completely known. So the function $h_{st}(\textbf{x})$ is an unknown function.

 Now we rewrite equation 23 as:
 \begin{equation}
 g_{st}(\textbf{x})= \Big[ h_{st}(\textbf{x}) \otimes \big[ h_{i}(\textbf{x}) \otimes I(\textbf{x}) \big] \Big] 
 \end{equation}
where $I(\textbf{x})$ is the fringe pattern in the case of monochromatic point source and $h_i(\textbf{x})$ denotes the inverse of convolution kernel $h_{re}(\textbf{x})$ shown in equation 19. Now using the associativity property of convolution, equation 25 becomes:
\begin{align}
g_{st}(\textbf{x}) &= \Big[ \big[ h_{st}(\textbf{x}) \otimes  h_{i}(\textbf{x}) \big] \otimes I(\textbf{x})  \Big] \\
&= \big[ h_{p}(\textbf{x})  \otimes I(\textbf{x}) \big] 
\end{align}
Equation 27 shows that the fringe pattern in case of a low spatio-temporal coherence light source is related to the fringe pattern corresponding to a monochromatic point source by a linear convolution operation. As mentioned for $h_{st}(\textbf{x})$, $h_p(\textbf{x})$ is again an unknown function, which we estimate using the imaging of a known object in our lensless imaging setup. The advantage of estimating $h_{p}(\textbf{x})$ instead of $h_{st}(\textbf{x})$ is that the number of unknown parameters to be estimated reduce drastically.

\begin{figure*}
\centering
{\includegraphics[width=17.25cm]{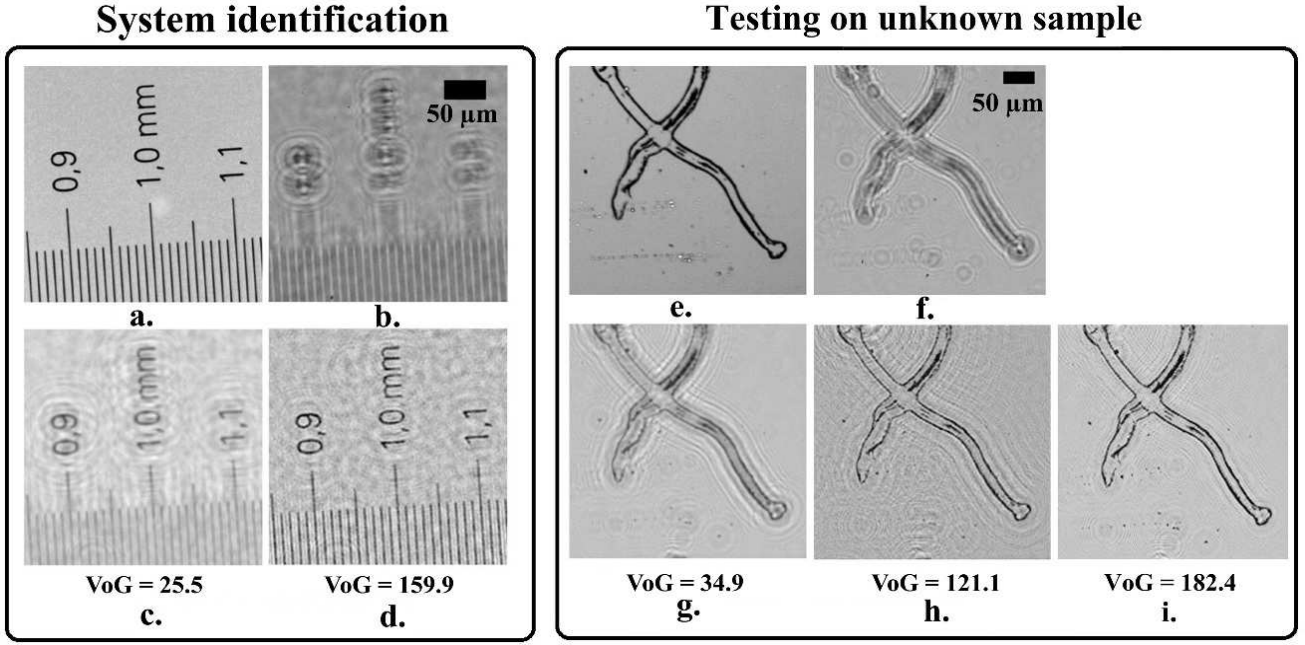}}
\caption{System identification:(a.) Leica's microscope calibration slide captured in a bright field microscope with 10x objective lens, used as ground truth, (b.) Fringe pattern for the same object recorded with the lensless microscopy setup with a spatially extended white LED, (c.) Reconstruction using angular spectrum method and (d.) Deconvolution using $h_{opt}$ shows improvement in variance of gradient (VoG); Testing: (e.) Photolithography sample captured in a bright field microscope with 10x objective lens, (f.) Fringe pattern for the same object recorded with the same lensless microscopy setup, (g.) Reconstruction using angular spectrum method, (h.) Deconvolution using $h_{opt}$ and (i.)Deconvolution using $h_{opt}$ with the upper-bound constraint.}
\label{Fig:4}
\end{figure*}
\begin{table*}
\caption{Quality measures of reconstructed images in figure 4g and 4i.}
\centering
\begin{tabular}{ccc}
\hline
Quality measure & Angular Spectrum method & Proposed method \\

\hline
Mean square error & $1.6 \times 10^3$ & $1.4 \times 10^3$ \\
Peak signal-to-noise ratio (in db) & 16.1  & 16.7 \\
Structural similarity index (SSIM)$^*$ & 0.52 & 0.59 \\
\hline

\end{tabular}

\footnotesize{$^*$ SSIM w.r.t. image captured in a bright field microscope with 10X objective lens.}\\

\end{table*}
\subsection{Imaging setup and data acquisition} 
A lensless imaging setup is prepared with a white LED with center wavelength $\lambda_{0}=550$ nm and line-width $\Delta \lambda = \sim 250$ nm as the light source (without any spatial or temporal filtering). An image sensor of pixel pitch 1.12 $\mu$m $\times$ 1.12 $\mu$m and physical size 3.68 mm $\times$ 2.76 mm is fixed at a distance of $\sim20$ cm from the light source. Sample is fixed on a micrometer stage at submillimeter distance from the image sensor (see fig. 1). 
Digits and bars (feature thickness $\sim 3$\ $\mu$m  and $\sim 1.7$\ $\mu$m) printed on Leica's microscope calibration slide are used as the known micro-objects (ground truth objects). Ground truth images are captured in a Leica's bright field microscope with 10x objective lens. Their corresponding fringe patterns are captured in the described lensless imaging setup. 12 different regions of this microscope calibration slide are used to prepare a set of fringe pattern-ground truth pair. Each image in the set is of digital resolution 256 x 256. Accurate registration of this pair is an essential step in this process and is obtained by the warping of ground truth images. Fringe patterns are first backprojected to their corresponding object using angular spectrum method with $H^{-1}_{OTF}$ for the wavelength 550 nm. The corresponding points between these highly blurred object reconstructions and high resolution ground truth images are manually selected, followed by the warping.

As mentioned earlier, $H_{OTF}$ is a function of sensor to object distance, $z$ which may not be accurately known. It is computationally obtained by comparing the sharpness of images (using variance of gradient) backprojected with $H^{-1}_{OTF}(z)$ for several values of $z$ in a iterative manner with step size $\Delta\ z=0.001$ mm. While imaging known object $z$ was 0.31 mm. 

Photolithography sample of features $\sim$3 $\mu$m and red blood cells of diameter $\sim$6-7 $\mu$m have been used as unknown objects. The object to sensor distance should be same as the value during the known sample imaging step. Since this may be difficult to achieve experimentally, distance becomes $z \pm \Delta z$. In our imaging experiment, $\Delta z$ was $\sim$-30 $\mu$m for photolithography sample and $\Delta z$ was $\sim$-10 $\mu$m for red blood cells.

\begin{figure*}
\centering
{\includegraphics[width=17cm]{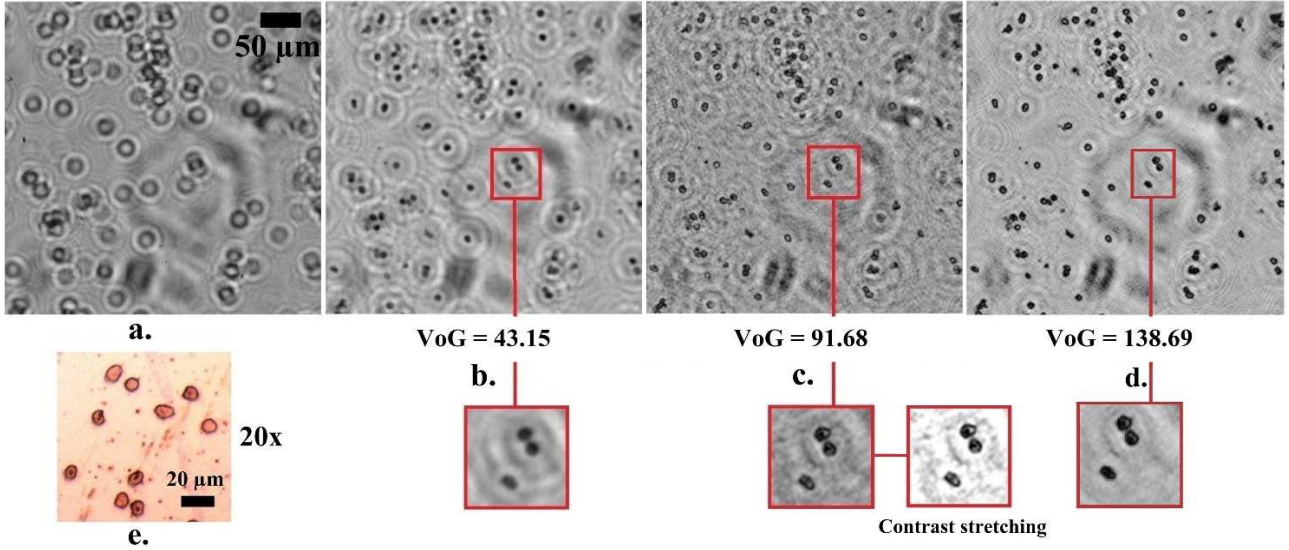}}
\caption{(a.) Fringe pattern for the Red blood cells’ of diameter $\sim$6-8 $\mu$m recorded with the lensless microscopy setup with a spatially extended white LED,(b.) Reconstruction using angular spectrum method, (c.) Deconvolution using $h_{opt}$, (d.) Deconvolution using $h_{opt}$ with upper-bound constraint and (e.) Red blood cells captured in a bright field microscope with 20x objective lens.}
\label{Fig:5}
\end{figure*}

\subsection{Estimation of $h_p$ via imaging a known sample}
$h_p(\textbf{x})$ is estimated by solving the following optimization problem:
\begin{equation}
\begin{split}
h_p(\textbf{x}) = \argmin_{\hat{h_{p}}(\textbf{x})} \ \sum_k \Big[ \ \lVert g^k_{st}(\textbf{x})- [ \hat{h_{p}}(\textbf{x}) \otimes I^{k}(\textbf{x}) ] \rVert \\ 
+ \alpha \lVert \hat{h_{p}}(\textbf{x}) \rVert \Big] \\
\textnormal{where} \ I^{k}(\textbf{x}) = h_{re}(\textbf{x},z) \otimes f^k(\textbf{x})
\end{split}
\end{equation}
$g^k_{st}(\textbf{x})$ and $f^k(\textbf{x})$ denote the $k$th fringe pattern and ground truth image from the set. $\alpha$ is the regularization parameter.  We obtained this minimization using interior point algorithm \cite{nocedal2006numerical, boyd2004convex, de2016limits}. The size of $h_p(\textbf{x})$ was empirically chosen to be 9 $\times$ 9. The size of $h_{re}(\textbf{x},z)$ was 128 $\times$ 128.
 
\subsection{Upper-bound constrained deconvolution for object estimation}
For computing an unknown object from the recorded fringe pattern, we have modified well known Richardson-Lucy deconvolution algorithm \cite{richardson1972bayesian,lucy1974iterative}. The original algorithm iteratively gives the maximum-likelihood object estimate for the data corrupted with poisson noise \cite{shepp1982maximum,dempster1977maximum}. The update rule is as following:
\begin{equation}
\hat{f}^{p+1}(\textbf{x}) = \hat{f}^{p}(\textbf{x}). \Bigg[ h_{opt}(-\textbf{x}) \otimes \frac{g_{st}(\textbf{x})}{h_{opt}(\textbf{x}) \otimes \hat{f}^{p}(\textbf{x})} \Bigg]
\end{equation}
where $p$ denotes the iteration number, and $h_{opt}(\textbf{x})$ is obtained by combining $h_p(\textbf{x})$ (estimated in previous section) and $h_{re}(\textbf{x},z\pm\Delta z)$ as:
\begin{equation}
h_{opt}(\textbf{x})= h_{p}(\textbf{x}) \otimes h_{re}(\textbf{x},z\pm\Delta z)
\end{equation}
All ones is used as the first estimate of $\hat{f}^p(\textbf{x})$. Without any regularization, this algorithm is strongly susceptible to noise overfitting problem. To avoid the noise overfitting and to suppress the ringing artifacts in the reconstruction, we have used the following additional constraint at every iteration:
\begin{align}
&\textnormal{If}\ \hat{f}^{p+1}(\textbf{x})> \tau \ \textnormal{UB}(\textbf{x}); \\
&\hat{f}^{p+1}(\textbf{x}) = \tau \ \textnormal{UB}(\textbf{x}) + \beta\ \Big(\hat{f}^{p+1}(\textbf{x}) - \tau \ \textnormal{UB}(\textbf{x}) \Big)
\end{align}
where $\beta$ is a parameter with value between 0 and 1, $\tau$ is a parameter close to 1 and $\textnormal{UB}(\textbf{x})$ is reference illumination, i.e. image without any object but with same illumination and exposure time of sensor.

\section{Results and discussion}
Figure 3 demonstrates the loss of resolution and object reconstruction quality in terms of artifacts when a light source of low spatio-temporal coherence is used for lensless microscopy. Both the images have been reconstructed with the angular spectrum method.
The resolution is degraded by atleast three to four times (when object to sensor distance is $\sim$300 $\mu$m; this fold of degradation of resolution is dependent on this distance). Figure 4a and 4b shows one of the 12 images from the set of ground truth-fringe pattern pair, used in section 2.4 for the estimation of $h_p(\textbf{x})$. Figure 4 and 5 demonstrates the improvement of the contrast, resolution and suppression of ringing artifacts using the proposed method. Table 1 shows a quantitative measure of the reconstructed image quality.
\begin{table*}

\centering

\caption{Different methods of lensless microscopy in low coherence light.}

\begin{tabular}{cccc}
\hline
Light source & Reconstruction method &  Reported Resolution & Speckle noise \\

\hline
\underline{Previously reported}: \\
470 nm LED, spatially extended & Feng and Wu, 2017  & 3.1 $\mu$m & No  \\
Direct sunlight & Feng and Wu, 2019a  & $\sim$6 $\mu$m & No  \\
Direct sunlight & Feng and Wu, 2019b  & 3.48 $\mu$m & No  \\
\underline{Our implementations}: \\
Laser, single mode optical fiber & Angular spectrum method  & $\sim$1-2 $\mu$m & Yes  \\
White LED, spatially extended & Angular spectrum method  & $\sim$4 $\mu$m & No  \\
White LED, spatially extended & Proposed method  & $\sim$1-2 $\mu$m & No  \\
\hline

\end{tabular}

\end{table*}

Variance of gradient (VoG) is a well-known measure of focus level because a well-focussed image is expected to have sharper edges \cite{pech2000diatom}. Similarly, a high optical resolution image reconstruction is expected to have sharper edges and hence we have used VoG as a quality measure of reconstructions. However VoG alone cannot be used as a resolution measure, a visual assessment of correct feature retrieval is essential. From the reconstruction of red blood cells using proposed method in figure 5c. and comparison with the gold standard images obtained using Leica's bright-field transmission-mode images, a resolution of $\sim$1-2 $\mu$m has been anticipated. The intracellular nucleus like feature of same size which were previously not resolvable (in figure 5b.) are well-resolved using the proposed method (in figure 5c.). Table 2 enlists the various different reconstruction methods with low coherence light sources and their reported resolution values.

\begin{figure}
\centering
{\includegraphics[width=8cm]{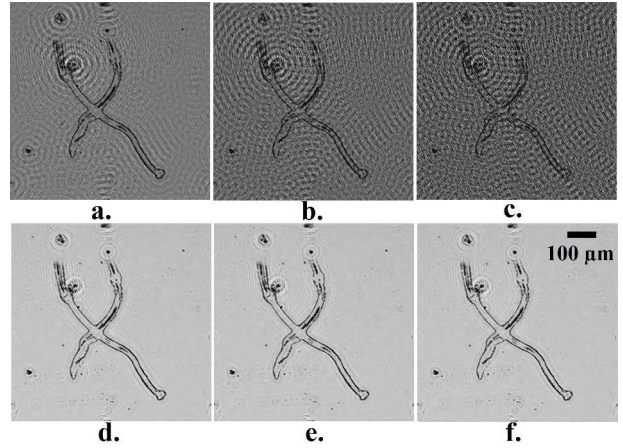}}
\caption{First row shows the reconstructions without upper-bound constraint after (a.) 15, (b.) 50 and (c.) 100 iterations; Second row shows the reconstructions with upper-bound constraint after (d.) 15, (e.) 50 and (f.) 100 iterations.}
\label{Fig:5}
\end{figure}

Figure 6(a-c) demonstrates that the non-regularized deconvolution algorithm suffers from the problems like noise overfitting and increasingly more and more pronounced ringing artifacts as the number of iterations increase. Figure 6(d-f) demonstrates the efficacy of upper-bound constraint to control these problems. As shown, proposed algorithm converges to an acceptable solution within 15 iterations. Reconstruction time taken for a 512 $\times$ 512 image was $\sim$1 second, computations being performed on MATLAB 2018b on a system with intel core i5-7500 CPU and 8 GB RAM.

\section{Conclusion}
In the previously established methods of lensless microscopy, backprojection is performed using the free space optical transfer function which is obtained from the scalar diffraction theory for monochromatic light. 
We have demonstrated that when a light source of low coherence like spatially extended white LED is used, this monochromatic free space transfer function does not give the best image reconstruction, specifically in terms of resolution and ringing artifacts. In the proposed method, using the imaging of known microscopic objects and principle of optimization, a suitable light-source-specific kernel has been estimated. Using this kernel and upper-bound constrained deconvolution algorithm, an image reconstruction with improved resolution (resolution $\sim$1-2 $\mu$m) and reduced ringing artifacts has been obtained.

\section{Acknowledgments}
Sanjeev Kumar acknowledges Council of Scientific and Industrial Research (CSIR), India (File No: 09/081(1282)/2016-EMR-1) for the award of an individual senior research fellowship. All the authors acknowledge Prof. Soumen Das, Ms. Jyotsana Priyadarshani and Mr. Prasoon Awasthi; Bio-MEMS lab, IIT-Kharagpur for helping with the imaging samples and the microscopy. All the authors thank the anonymous reviewers.

\bibliographystyle{cas-model2-names}
\bibliography{Sanjeev_OLE_2019}

\end{document}